\documentclass[%
 aip,
rsi,%
 amsmath,amssymb,
 reprint,%
]{revtex4-1}

\usepackage{graphicx}% Include figure files
\usepackage{dcolumn}% Align table columns on decimal point
\usepackage{bm}% bold math
\begin{document}

\preprint{AIP/123-QED}

%\title[Sample title]{Resonance control of graphene drum resonator in nonlinear regime by standing wave of light}
\title[]{Resonance control of graphene drum resonator in nonlinear regime by standing wave of light}
% Force line breaks with \\
%\thanks{Footnote to title of article.}

\author{Taichi Inoue}
\author{Yuki Anno}
\author{Yuki Imakita}
\author{Kuniharu Takei}
\author{Takayuki Arie}
\author{Seiji Akita}
 \email{akita@pe.osakafu-u.ac.jp}
\affiliation{Department of Physics and Electronics, Osaka Prefecture University, Sakai 599-8531, Japan}

%\author{A. Author}
% \altaffiliation[Also at ]{Physics Department, XYZ University.}%Lines break automatically or can be forced with \\
%\author{B. Author}%

%\affiliation{ Authors' institution and/or address%\\This line break forced with \textbackslash\textbackslash}%

%\author{C. Author}
% \homepage{http://www.Second.institution.edu/~Charlie.Author.}
%\affiliation{Second institution and/or address%\\This line break forced% with \\}%

\date{\today}% It is always \today, today,
             %  but any date may be explicitly specified

\begin{abstract}
We demonstrate the control of resonance characteristics of a drum type graphene mechanical resonator in nonlinear oscillation regime by the photothermal effect, which is induced by a standing wave of light between a graphene and a substrate. Unlike the conventional Duffing type nonlinearity, the resonance characteristics in nonlinear oscillation regime is modulated by the standing wave of light despite a small variation amplitude. From numerical calculations with a combination of equations of heat and motion with Duffing type nonlinearity, this can be explained that the photothermal effect causes delayed modulation of stress or tension of the graphene. 
%
%Valid PACS numbers may be entered using the \verb+\pacs{#1}+ command.
\end{abstract}

%\pacs{Valid PACS appear here}% PACS, the Physics and Astronomy
                             % Classification Scheme.
%\keywords{Suggested keywords}%Use showkeys class option if keyword
                              %display desired
\maketitle

%\begin{quotation}
%The ``lead paragraph'' is encapsulated with the \LaTeX\ 
%\verb+quotation+ environment and is formatted as a single paragraph before the first section heading. 
%(The \verb+quotation+ environment reverts to its usual meaning after the first sectioning command.) 
%Note that numbered references are allowed in the lead paragraph.
%
%The lead paragraph will only be found in an article being prepared for the journal \textit{Chaos}.
%\end{quotation}

%\section{\label{sec:level1}First-level heading:\protect\\ The line break was forced \lowercase{via} \textbackslash\textbackslash}

A graphene has attracted much attention as a component of nano-electro-mechanical systems (NEMS) because of its superb mechanical and electrical properties. Especially, in terms of the graphene NEMS, a graphene mechanical resonator (G-MR)\cite{Bunch2007, Zande2010, Barton2012, Croy2012, Jiang2012, Miao2014, Chen2016, Yuasa2011, Takamura2014, Takamura2013} is expected to be a highly sensitive mass and force sensor and a component of the computing application. To improve the sensitivity, it is important to realize the nanomechanical resonator with high quality factor (Q-factor). Intrinsic Q-factor depends only on the material itself, so that the modulation of intrinsic Q-factor is not easy. Contrary, the apparent Q-factor can be easily modulated by modulating the loss part of the motion of the resonator by applying the active \cite{Humphris2000, Horber2003} or passive feedback \cite{Metzger2008, Yasuda2016, Metzger2004} to the resonating system  under linear response regime. In the case of passive feedback, the photothermal self-oscillation and laser cooling of G-MR with a Fabry-Perot cavity between graphene and substrate have been demonstrated by photothermal back-action induced by positive and negative feedback under linear oscillation regime, respectively\cite{Barton2012}. This photothermal back-action was analyzed based on the delayed response component in the resonating system\cite{Yuasa2011, Metzger2004}. In addition to linear oscillation, nonlinear effects on mechanical resonators, which are commonly described by the Duffing effect at large oscillation amplitude, are important for the improvement of sensitivity or other applications such as mode coupling toward computing applications \cite{Miao2014, Chen2016, Bagheri2011, Isacsson2007, Nagataki2013, Kagota2013}. However, there is limited number of reports for the resonance control of G-MR in nonlinear resonance regime with the delayed component using optical back-action\cite{Croy2012, Miao2014}. In addition, the delayed component using optical back-action with steep optical field gradient should have a position dependence. In this case, we have to analyze the combination of the Duffing type nonlinear equation and the delayed-time component with position dependence. However, this type of equations is hard to obtain the accurate analytical solution. Thus, the detailed analysis of the nonlinear resonance and control of its resonance properties of G-MR in nonlinear regime are still challenging task. In this study, we demonstrate the resonance control of the G-MR in the nonlinear resonance regime with a Fabry-Perot cavity using the optical back-action induced by the standing wave of light in the Fabry-Perot cavity. In addition to the experiment, we perform numerical calculations to investigate the relation between the nonlinearity and the position dependent delayed-time component induced by the optical back-action in the nonlinear oscillation regime.
%The Introduction section, of referenced text\cite{Figueredo:2009dg} expands on the background of the work (some overlap with the Abstract is acceptable). The introduction should not include subheadings.

\begin{figure}[ht]
\centering
\includegraphics[width=60mm]{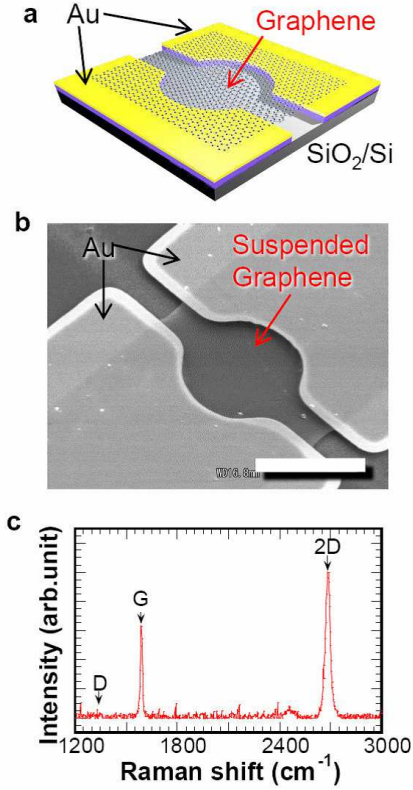}
\caption{{\bf Drum type graphene mechanical resonator with a Fabry-Perot cavity.} (a) Schematic illustration of the drum type G-MR with a Fabry-Perot cavity. (b) SEM image of the drum type G-MR with the drum diameter of 5.4 $\mu$m. The scale bar is 5 $\mu$m. (c) Raman spectrum for whole of the suspended part of the drum type G-MR.}
\label{fig1}
\end{figure}

%\section*{Results}
A drum type G-MR as schematically shown in Fig. \,\ref{fig1}(a) was investigated in this study. The fabrication process is as follows as schematically illustrated in Fig. S1 of ``{\it Supplementary information}''. First, metal electrodes consisting of Cr/Au (5 nm/30 nm) as a support of graphene drum were fabricated on a heavily doped Si substrate with a 300 nm-thick SiO$_2$ layer by conventional photolithography process (Fig. S1(b)). Subsequently, a monolayer graphene was transferred onto the substrate by using polymethyl methacrylate (Fig. S1(c)) and was trimmed using oxygen plasma etching to form the drum type G-MR (Fig. S1(d)), where the graphene was synthesized using low-pressure chemical vapor deposition at 1000 $^\circ$C using Cu foil as catalyst \cite{Anno2014, Li2009}. To form the drum type G-MR suspended by metal electrodes, the SiO$_2$ layer underneath the graphene drum was etched by buffered HF (Fig. S1(e)), where the metal electrodes are used as the metal mask for etching. Note that the trenches at both sides of the drum were required for uniform etching of SiO$_2$ layer and acts as the evacuation channel of air between the graphene drum and the substrate during the measurement of resonance characteristics. The samples thus fabricated were finally dried by using supercritical drying to prevent sticking of suspended graphene induced by surface tension of water. As shown in Fig. \,\ref{fig1}(b), the drum type G-MR with a diameter of 5.4 $\mu$m was successfully fabricated. To investigate the degradation of graphene after the fabrication process, the Raman spectroscopy was performed. The D band originated from the defects on Raman spectrum at suspended part of the G-MR is rarely observed as shown in Fig. \,\ref{fig1}(c). This implies that this fabrication process induces no significant degradation of the graphene.

\begin{figure}[ht]
\centering
\includegraphics[width=60mm]{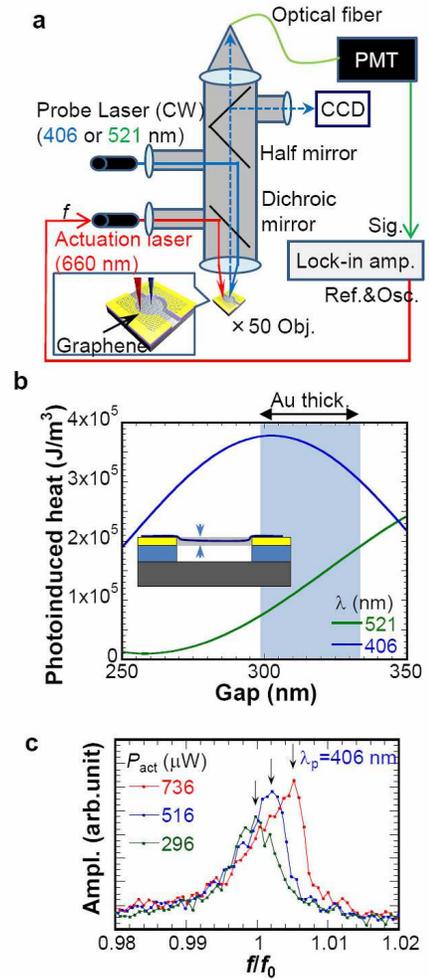}
\caption{{\bf Measurement setup for resonance characteristics with probe lasers.} (a) Schematic illustration of the optical detection and actuation system. (b) Position dependence of the heat induced by the interference of probe laser (406 or 521 nm) between the substrate surface and the graphene membrane obtained from FEM analysis. (c) Resonance curve measured under various actuation laser intensities, where the wavelength and the intensity of the probe laser are 406 nm and 6.3 $\mu$W, respectively. The frequency $f$ is normalized by the resonance frequency $f_0$ ($\sim$ 9.83 MHz) at linear oscillation.}
\label{fig2}
\end{figure}

The resonance of the drum type G-MR was measured by using optical detection method in vacuum at $\sim$ 10$^{-3}$ Pa as shown in Fig. \,\ref{fig2}(a). The light interference intensity induced between the G-MR and Si substrate was measured to detect the displacement of the G-MR, where a laser with wavelengths of 406 or 521 nm was used as a probe, which irradiated the center of the drum. To oscillate the G-MR, a laser of 660 nm wavelength was irradiated around the support edge of the G-MR with a spot diameter of $\sim$ 1 $\mu$m as schematically illustrated in Fig. \,\ref{fig2}(a), which was modulated at a certain frequency, $f$, through an objective lens with a numerical aperture (NA) of 0.7.

Figure \,\ref{fig2}(b) shows the graphene position dependence of the induced heat on G-MR by the interfered probe laser light between the graphene and Si substrate, which were calculated by finite element method (FEM), where the complex refractive indexes of graphene (2.6-1.3i) \cite{Blake2007} and Si (5.6-0.4i at 400 nm) \cite{Palik1998} were used. The gray area between 300 and 335 nm corresponds to the graphene height from the substrate deduced from the SiO$_2$ and electrodes thickness as schematically illustrated in the inset of Fig. \,\ref{fig2}(b), where there is a possibility that the edge of graphene partially sticks to the side of Au electrodes. Note that an accurate height of the graphene was hard to measure using atomic force microscopy (AFM) because of insufficient tension of the graphene, which caused the unexpected perturbation on the AFM image. In the gray area, the slopes of the profiles for the wavelengths $\lambda_p$ of 406 and 521 nm were opposite. This difference cause the different response of the photothermal effect induced by the probe laser on the drum type G-MR for respective $\lambda_p$. The thermal expansion coefficient of the graphene shows negative (-8$\times$10$^{-6}$ K$^{-1}$) around the room temperature \cite{Yoon2011}, so that the higher heat generation at graphene induces the higher stress \cite{Takamura2014} or tension of the graphene membrane.

Figure \,\ref{fig2}(c) shows one of example for the frequency response curve of the drum type G-MR taken from lower frequency ($f_0$ $\sim$ 9.83 MHz at the resonance of linear oscillation regime), where the actuation laser intensity ($P_{act}$) was changed from 296 to 736 $\mu$W with $\lambda_p$ = 406 nm. All of the frequency response curves shown in this paper were taken from lower frequency. At higher intensity of the actuation, the hardening type nonlinear vibration is clearly observed, where the resonance curve deforms from symmetrical to unsymmetrical and resonance peak shifts toward higher frequency with increasing the actuation.

The nonlinear vibration observed here could be excited by increasing amplitude which led to the appearance of the nonlinearity term in resilience. In this case, the Duffing type equation of motion with external force $F$cos$\omega t$ is applicable with the amplitude $x$ described by
\begin{equation}
m\ddot{x}+\gamma\dot{x}+k_0x+\beta x^3=F \cos \omega t,
\end{equation}
where $m$ is the mass of G-MR, $\gamma$ is linear damping related to quality factor (Q-factor), $k_0$ is the intrinsic spring constant, $\beta$ is the nonlinear coefficient called Duffing constant, $\omega$ is angular frequency. The spring constant of the drum type G-MR is determined from the stress or tension acting on the graphene. At small amplitude, $k_0$ is deduced to be constant, while, under larger amplitude, the deflection of graphene induces additional hardening like $k_0+\Delta k$, which results in the nonlinear vibration. We have confirmed that the resonance properties under the linear regime showed no obvious dependence of $P_p$ (from 6.30 to 9.28 $\mu$W) as shown in Fig. S2 of ``{\it Supplementary information}''. This implies that no significant modulations of $\gamma$ and $k_0$ under the linear regime are induced by the photo-thermal effect due to the standing wave of the probe lasers.

\begin{figure}[ht]
\centering
\includegraphics[width=60mm]{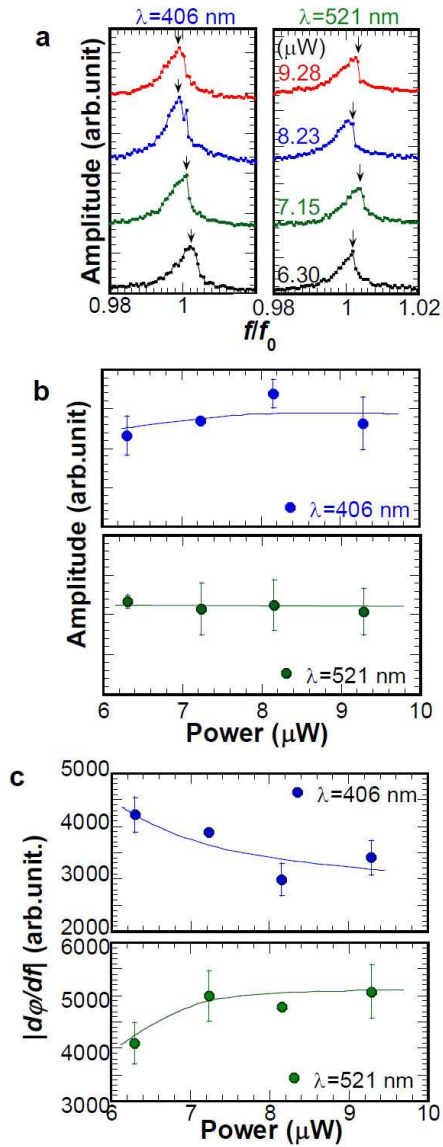}
\caption{{\bf Nonlinearity control of the drum type G-MR by interference of light.} (a) Frequency response curves with various probe laser intensities $P_p$ measured by $\lambda_p$= 406 and 521 nm under $P_{act}$ = 516 $\mu$W. Arrows indicate respective resonances, where the frequency $f$ is normalized by the resonance frequency at linear response measured under weak actuation. (b) Probe laser intensity $P_p$ dependence of the amplitude at resonance shown in (a). (c) Probe laser intensity $P_p$ dependence of nonlinearity determined by $d\varphi/df$. Solid lines are guide for eyes.}
\label{fig3}
\end{figure}

In order to modify the resonance characteristics of the G-MR in nonlinear oscillation regime using the interference light, the probe laser intensity ($P_p$) dependence was investigated at $P_{act}$ = 516 $\mu$W corresponding to the weak nonlinear oscillation shown in Fig. \,\ref{fig2}(c). Figure \,\ref{fig3}(a) shows the frequency response curves obtained at various probe laser intensities from 6.30 to 9.28 $\mu$W for respective $\lambda_p$. The frequency responses show the weak nonlinearity for both $\lambda_p$. In order for detailed analysis of the nonlinearity, we evaluated slopes of the frequency response of the phase shift $d\varphi/df$ at the resonance as the measure of nonlinearity, where $\varphi$ is the phase shift between the deflection of G-MR and the actuation force. Briefly, at linear oscillation regime, $d\varphi/df$ is determined by $2m/\gamma$, which should be independent of $\lambda_p$. In the case of nonlinear oscillation regime, the frequency response of the phase shift shows abrupt change from positive to negative so called jumping phenomenon, so that $d\varphi/df$ at resonance is infinity. Consequently, one can expect that the weak nonlinear condition corresponding to the intermediate region gives a certain slope $d\varphi/df$ depending on the nonlinearity. Note that, in case of the conventional nonlinear oscillation governed by Eq. 1, the frequency at the resonance shifts toward higher frequency with increasing the amplitude. In addition, the nonlinearity also increases with increasing the amplitude.

Figures \,\ref{fig3}(b) and \,\ref{fig3}(c) show the probe laser intensity dependence of the maximum amplitude and nonlinearity $d\varphi/df$ at resonance. In the case of $\lambda_p$ = 406 nm, the $d\varphi/df$ at resonance decreased despite the slight increase in amplitude with increasing $P_p$. In addition, as clearly observed in Fig. \,\ref{fig3}(a), the peak position for $\lambda_p$ = 406 nm shifts 0.4\% toward lower frequency with increasing $P_p$. These change induced by the probe laser is completely opposite for the conventional nonlinear phenomenon explained by Eq. 1. In the case of $\lambda_p$ = 521 nm, the amplitude at resonance shows no significant probe laser intensity dependence. The peak position at resonance is fluctuated within 0.3\% and shows no specific dependence. Only the nonlinearity $d\varphi/df$ increases with increasing Pp., which is the opposite dependence observed at $\lambda_p$ = 406 nm, which may come from the opposite slopes of the profiles of the standing wave of the probe laser shown in Fig. \,\ref{fig2}(b). The changes also cannot be explained by the conventional Duffing equation described by Eq. 1. It is noteworthy that the resonance characteristics in the linear regime shows no significant change for changing the probe laser intensity as mentioned above. Thus, we can conclude that the nonlinearity is greatly modified by the standing wave of light.

To investigate the unexpected resonance properties in nonlinear regime, we consider the delayed time response, which was considered as the origin of the modification of the resonance properties in linear resonance regime \cite{Metzger2008}. In the case of linear resonance regime, the delayed time response on the oscillating system acts only on the phase component, which results in the modulation of damping, $\gamma$ \cite{Metzger2008, Yasuda2016}. Similar delay component induced by the photothermal effect of the standing wave of light should be considered for the G-MR under nonlinear regime. The temperature change of G-MR, $\Delta \Theta$ modifies the stress or tension acting on the graphene. Note that $\Delta \Theta$ depends not only on the position $x$ but also time $t$ due to the thermal relaxation time $\tau_{th}$ determined by the heat capacity and the thermal resistance, where the heat capacity considered here should include not only the graphene itself but also the substrate contacted to the graphene. In the case of nonlinear oscillator with the time delayed component, the analytical solution is hard to obtain. 

\begin{figure}[ht]
\centering
\includegraphics[width=60mm]{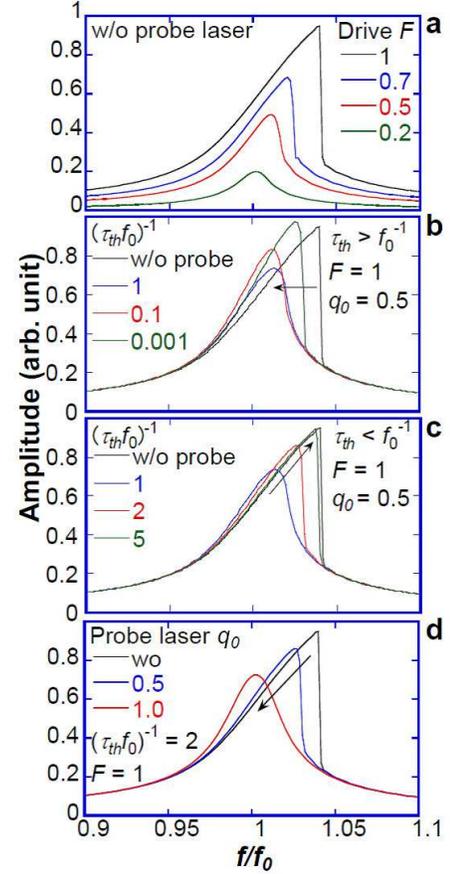}
\caption{{\bf Drum type graphene mechanical resonator with a Fabry-Perot cavity.} (a) Schematic illustration of the drum type G-MR with a Fabry-Perot cavity. (b) SEM image of the drum type G-MR with the drum diameter of 5.4 $\mu$m. The scale bar is 5 $\mu$m. (c) Raman spectrum for whole of the suspended part of the drum type G-MRNumerical calculation of the Duffing type nonlinear resonance with delayed effect. (a) Conventional Duffing resonance curves with various driving amplitude. (b) and (c) Frequency response curves calculated with various delayed time constant, $\tau_{th}$, for $\tau_{th} > {f_0}^{-1}$ and $\tau_{th} < {f_0}^{-1}$, respectively. (d) Frequency response curves calculated with various probe laser intensity $P_p$.}
\label{fig4}
\end{figure}

Assuming the linear relation between the spring constant and temperature, the additional spring constant $\Delta k$ is given by
\begin{equation}
\Delta k=-\alpha \Delta \Theta ,
\end{equation}
where $\alpha$ is the thermal expansion coefficient of graphene. Additionally, the position dependence of the probe laser induced heat is approximated to be linear because of the small oscillation amplitude in comparison of the gap between the graphene and the substrate. We further approximate that this simplified model with a lumped capacitance model for heat equation, $\Delta \Theta$ is applicable expressed by
\begin{equation}
\frac{d\Delta \Theta}{dt}=q_0x-\frac{\Delta \Theta}{\tau_{th}},
\end{equation}
where $q_0$ is a slope of the position dependence of the induced heat normalized by the system heat capacity on graphene, which is defined by $P_p$ and profiles of photogenerated heat as shown in Fig. \,\ref{fig2}(b). This expression is similar to the case described in Ref. 13 for linear oscillation except the position dependence term, $q_0x(t)$, proposed here. Thus, the additional spring constant, $\Delta k$ is expected to be a function with nonlinear dependence on $t$ and $x$ as $\Delta k(x, t)$. In this case, the Duffing type motion of equation described by Eq. 1 is now modified to be 
\begin{equation}
m\ddot{x}+\gamma\dot{x}+(k_0+\Delta k)x+\beta x^3=F \cos \omega t,
\end{equation}
where the term $\Delta k(x, t)$ consists of higher order terms of $x$ with certain phase shifts resulting from the time delayed component, which gives raise to the modification of $\gamma$ and $\beta$ in addition to the spring constant itself. Solving simultaneous differential Eqs. 2-4 numerically, one can {\it qualitatively} evaluate the photothermal effect induced by the probe laser. Parameters used in this calculation are listed in Table I. As a reference, the nonlinear behavior of this model was examined as shown in Fig. \,\ref{fig4}(a), where the driving amplitude $F$ was changed without the photothermal effect induced by the probe laser. A linear response can be obtained at $F$ = 0.2. The nonlinearity with apparent hardening effect increases with increasing the driving amplitude and the jumping phenomenon on the frequency response curve is clearly observed. 

\begin{table}[ht]
\centering
\caption{\label{tab1}Parameters used in the numerical calculation.}
\begin{tabular}{cccc}
\hline
$(k_0/m)^{0.5}$ & $\gamma$ & $\beta$ & $\alpha$\\
(rad. s$^{-1}$)	 & (N s m$^{-1}$) & (N m$^{-3}$) & (N m$^{-1}$ K$^{-1}$) \\
%Condition & n & p \\
\hline
1 & 1/50 & 0.5 & -2\\
\hline
\end{tabular}
\end{table}

To evaluate the effect of the delayed time constant $\tau_{th}$ on the nonlinearity, where $\tau_{th}f_0$ is used as a parameter \cite{Metzger2008}, which is a ratio of the oscillation period at linear oscillation, ${f_0}^{-1}$, and the delayed time constant $\tau_{th}$. In the case of $(\tau_{th}f_0)^{-1} < 1$ corresponding to slow response of temperature, the nonlinearity decreases with increasing $(\tau_{th}f_0)^{-1}$ as observed in Fig. \,\ref{fig4}(b). The temperature modulation induced by the oscillation is decreased and finally becomes zero at longer $\tau_{th}$, which results in the elimination of the delayed effect. At $(\tau_{th}f_0)^{-1} > 1$ corresponding to faster response of temperature to the oscillation period, the nonlinearity increases with increasing $(\tau_{th}f_0)^{-1}$. This indicates that fast response, i.e., no delay effect, gives rise to no effect on the phase component of the resonator. Thus, the nonlinearity returns to original state with increasing $(\tau_{th}f_0)^{-1}$. Thus, the delayed time response greatly influences the resonance characteristics and the condition of $(\tau_{th}f_0)^{-1} \approx 1$ is the most efficient to suppress the nonlinearity of the resonator, which is very similar to the case for laser cooling of the mechanical resonator \cite{Yuasa2011}.

Figure \,\ref{fig4}(d) represents the light intensity of the probe laser dependence of the frequency response at $(\tau_{th}f_0)^{-1} = 2$. The nonlinearity decreases with increasing the light intensity and vanished at $P_p$ = 1. Note that the amplitude at the resonance is $\sim$ 0.7, which is much larger than 0.2 obtained in the case of no delay effect shown in Fig. \,\ref{fig4}(a). Further increase of $P_p$ induces the softening effect on the resonance characteristics. These indicate that the delay effect does not merely suppress nonlinearity by decreasing the amplitude as the conventional nonlinear oscillator, but directly reduces the nonlinear term in Eq. 4. This weak amplitude dependence of the nonlinearity is consistent with the experimental results of weak amplitude dependence of the nonlinearity as shown in Fig. \,\ref{fig2}. To perform the quantitative analysis, we have to know the total heat capacity of the drum type G-MR and thermal expansion coefficient of the CVD grown graphene which includes the grain boundary. These are subject for further study.

We demonstrated the control of nonlinearity of the drum type G-MR by the photothermal effect, which is induced by the standing wave of light between graphene and substrate namely a Fabry-Perot cavity. Unlike the conventional Duffing type nonlinearity, the nonlinearity was modulated with slight oscillation amplitude variation by changing the wavelength and intensity of the probe laser. This phenomenon can be explained that the delayed thermal response induced by the photothermal effect causes the modulation of the effective spring constant of the graphene membrane in phase space. We performed numerical calculations consisting of the heat equation and the Duffing type equation of motion with the delayed-time response to investigate the modulation of resonance characteristics in nonlinear oscillation regime. As a result, the weak amplitude dependence of the nonlinearity observed in the experimental results was qualitatively consistent with the calculations. We believe that this modification of oscillation characteristics in nonlinear regime opens the way to realize NEMS circuit using the nonlinear effect such as up-conversion of the frequency or mode coupling.

\section*{Acknowledgements}

This work was partially supported by JSPS KAKENHI Grant Numbers JP15H05869, JP16K14259, JP16H00920, JP17H01040 and JP16H06504.

%\nocite{*}
\bibliography{sample}

\end{document}